\documentclass[
reprint,
amsmath,amssymb,
aps,
]{revtex4-2}

\usepackage{graphicx}
\usepackage{dcolumn}
\usepackage{bm}
\usepackage[colorlinks, linkcolor=blue, anchorcolor=blue, citecolor=blue]{hyperref}
\usepackage{url}
\usepackage{amsmath,amssymb}
\usepackage[mathlines]{lineno}
\usepackage{subfigure}
\usepackage{eurosym}
\usepackage{amsmath,amssymb,mathrsfs}

\begin{document}

\preprint{APS/123-QED} 

\title{Simulating and probing many-body quantum states in waveguide-QED systems with giant atoms}
\author{C. L. Yang}
\affiliation{School of Physical Science and Technology, Southwest Jiaotong University, Chengdu 610031, China}
\author{W. Z. Jia}
\email{wenzjia@swjtu.edu.cn}
\affiliation{School of Physical Science and Technology, Southwest Jiaotong University, Chengdu 610031, China}

\date{\today}

\begin{abstract}
Waveguide quantum electrodynamics (wQED) with giant atoms provides a distinctive opportunity to quantum simulation of many-body physics through its unique decoherence-free interactions. This study presents a theoretical framework for simulating the diagonal Aubry-Andr\'e-Harper (AAH) model in the context of giant-atom wQED. The proposed scheme employs photonic modes in the waveguide to not only mediate interactions between atoms but also to detect the energy spectrum of the atom array. To illustrate the effectiveness of this approach, we present a simulation of the Hofstadter butterfly spectrum with high precision. Furthermore, for an incommensurate AAH atomic chain, we demonstrate that the photon transmission spectrum can accurately distinguish between the many-body localized phase and the extended phase.
The method presented here is also applicable to the simulation of other types of 1D atomic chains based on giant-atom wQED.
\end{abstract}

\maketitle


\section{\label{introduction}INTRODUCTION}
Waveguide quantum electrodynamics (wQED) systems \cite{Roy-RMP2017, Gu-PhysReports2017,Sheremet-RMP2023} with giant atoms \cite{Kockum-MI2021} have emerged as a novel paradigm in quantum optics. In these systems, quantum systems are coupled to distant positions of the waveguide, which are separated by wavelength spacings. The structures of giant atoms have been realized in a variety of quantum systems, including superconducting qubits \cite{Andersson-NatPhy2019, Kannan-Nature2020} and giant spin ensembles \cite{Wang-Natcom2022}. Moreover, theoretical proposals have been put forth suggesting the use of cold atoms in optical lattices \cite{Tudela-PRL2019} and synthetic dimensions \cite{Du-PRL2022,Xiao-npj2022} as a means of realizing the giant-atom configurations. It is evident that giant atoms cannot be regarded as pointlike particles, and thus the conventional dipole approximation is no longer applicable. The non-dipole effects of giant atom systems can produce a number of novel phenomena, including frequency-dependent decay rate and Lamb shift \cite{Kockum-PRA2014}, non-Markovian dynamics \cite{Guo-PRA2017,Andersson-NatPhy2019,Guo-PRR2020,Qiu-SciChina2023,Lim-PRA2023,Xu-NewJPhys2024,Li-PRA2024,Roccati-PRL2024}, generation of enhanced entanglement \cite{Santos-PRL2023,Yin-PRA2023}, tunable atom-photon bound states \cite{Guo-PRA2020,Zhao-PRA2020,Wang-PRL2021,Cheng-PRA2022,Zhang-PRA2023,Jia-OE2024} and scattering states \cite{Ask-ArXiv2020,Cai-PRA2021,Feng-PRA2021,Zhu-PRA2022,Yin-PRA2022,Peng-PRA2023,Gu-PRA2023,Gu-PRA2024}. One of the most unique properties of giant-atom wQED is the decoherence-free interaction between a pair of braided giant atoms coupling to a linear waveguide \cite{Kockum-PRL2018,Carollo-PRR2020}, which has been successfully implemented in superconducting circuits \cite{Kannan-Nature2020}. This type of protected exchange interactions between giant atoms can be used to perform coherent quantum operations \cite{Du-PRA2023,Soro-PRA2023}.
Another promising application of such decoherence-free interactions is quantum simulation of coupled spins, such as one-dimensional (1D) 
tight-binding atomic chain \cite{Kockum-PRL2018}, 1D Su-Schrieffer-Heeger atomic chain \cite{Peng-PRA2023}, and driven-dissipative spin chains \cite{Chen-ArXiv2024}.

Quantum simulation provides a robust tool for investigating many-body phenomena that may present a significant challenge for classical computers \cite{Feynman-IJTP1982,Georgescu-RModPhys2014}. The 1D Aubry-Andr\'e-Harper (AAH) model \cite{Harper-PPSSA1955,Aubry-IPS1980}, as a prominent tool for investigating topological physics and localization transition, has attracted considerable attention from theoretical perspectives \cite{Simon-AAM1982,Thouless-PRB1983,Ostlund-PRL1983,Hiramoto-PRL1989,Jitomirskaya-JSTOR1999,Iyer-PRB2013,Ganeshan-PRL2013,Lellouch-PRA2014,Roosz-PRB2014,Liu-PRB2015,Khemani-PRL2017}. 
To illustrate, the diagonal AAH model can be precisely mapped to the two-dimensional (2D) Hofstadter model \cite{Hofstadter-PRB1976}, demonstrating a 2D quantum Hall effect (QHE) with topologically protected edge states. The corresponding energy spectra versus the dimensionless magnetic flux form a famous fractal structure known as the Hofstadter butterfly. In addition, the incommensurate diagonal AAH model describes a 1D tight-binding lattice with quasiperiodic potential, in which the disorder strength can induce a delocalization-localization
phase transition \cite{Aubry-IPS1980,Simon-AAM1982,Jitomirskaya-JSTOR1999}. 
Experimentally, the AAH model was first implemented on quantum simulation platforms based on cold atoms in optical lattices \cite{Roati-Nature2008,Michael-science2015,Bordia-Natphys2017} and photonic lattice \cite{Lahini-PRL2009,Kraus-PRL2012}. 
More recently, with the development of modern nanotechnology, superconducting quantum circuits have proven to be an optimal platform for verifying the underlying physics of the AAH model \cite{Roushan-science2017,Shi-PRL2023,Li-npjQI2023,Li-Natcom2023}.

The inherent compatibility between giant-atom wQED systems and superconducting quantum circuits \cite{Andersson-NatPhy2019,Kannan-Nature2020} motivates us to investigate the feasibility of simulating the AAH model based on giant-atom wQED.
Therefore, in this paper, we construct the diagonal AAH model by employing the decoherence-free interactions that are unique to giant-atom wQED systems. In our proposed scheme, photonic modes in 1D waveguide serve not only to mediate interactions between atoms, but also to facilitate the readout of the spectral structure of atomic chain. By appropriately modulating the coupling configurations between the atoms and the waveguide, it is possible to reproduce the energy spectrum of the atomic chain with high precision through the photon scattering process. As an example, we demonstrate the efficacy of this method by simulating the Hofstadter butterfly spectrum at high resolution.
Moreover, we find that for an incommensurate AAH atomic chain in the extended phase, the decays of the collective modes into the waveguide are concentrated on a few of these modes. In contrast, in the localized phase, the decays are distributed evenly in each mode. This interesting correspondence makes the localization transition easy to detect from the single-photon scattering spectrum.
These results also indicate that wQED with giant atoms is optimal for investigating light-matter interactions between an AAH atomic chain and its surrounding photonic environment. The methodology presented here can be extended to simulate other types of 1D atomic chains.

The remainder of this paper is organized as follows. In Sec.~\ref{TheoreticalDescription}, we derive the single-photon scattering amplitudes for wQED systems with multiple giant atoms based on the real-space formalism \cite{Shen-OL2005,Shen-PRL2005}. Additionally, we extract the effective Hamiltonian that describes the interacting atomic chain. In Sec.~\ref{SimulationAAH}, we discuss quantum simulation of AAH model through wQED systems with giant atoms. Specifically, we present a method for constructing an AAH-type atomic chain based on giant-atom wQED structure in Sec.~\ref{AAHscheme}. Subsequently, we provide discussions on probing the energy spectrum, simulating the Hofstadter butterfly, and probing the localization transition through single-photon scattering spectra in Sec.~\ref{ProbeEnergyBand}-\ref{LocalizationTransition}. Additionally, we discuss the influence of spontaneous emission to the non-waveguide degrees of freedom in Sec.~\ref{DissipationUnguided}.
Finally, further discussions and conclusions are given in Sec.~\ref{conclusion}.
\section{\label{TheoreticalDescription}Theoretical description of single-photon scattering in waveguide QED containing multiple giant atoms}
In this section, we provide an overview of the single-photon scattering problem for wQED with multiple giant atoms. From the derivation of the single-photon scattering amplitudes, we can extract the effective Hamiltonian describing the inter-atomic interactions mediated by the waveguide modes. This serves as the foundation for the construction of the target Hamiltonian that needs to be simulated. Additionally, in our proposal, single-photon scattering spectrum is also employed as a means to probe the many-body states of an atomic chain.
\subsection{\label{Hamiltonian}Hamiltonian and equations of motion}
\begin{figure}[t]
	\centering
	\includegraphics[width=0.5\textwidth]{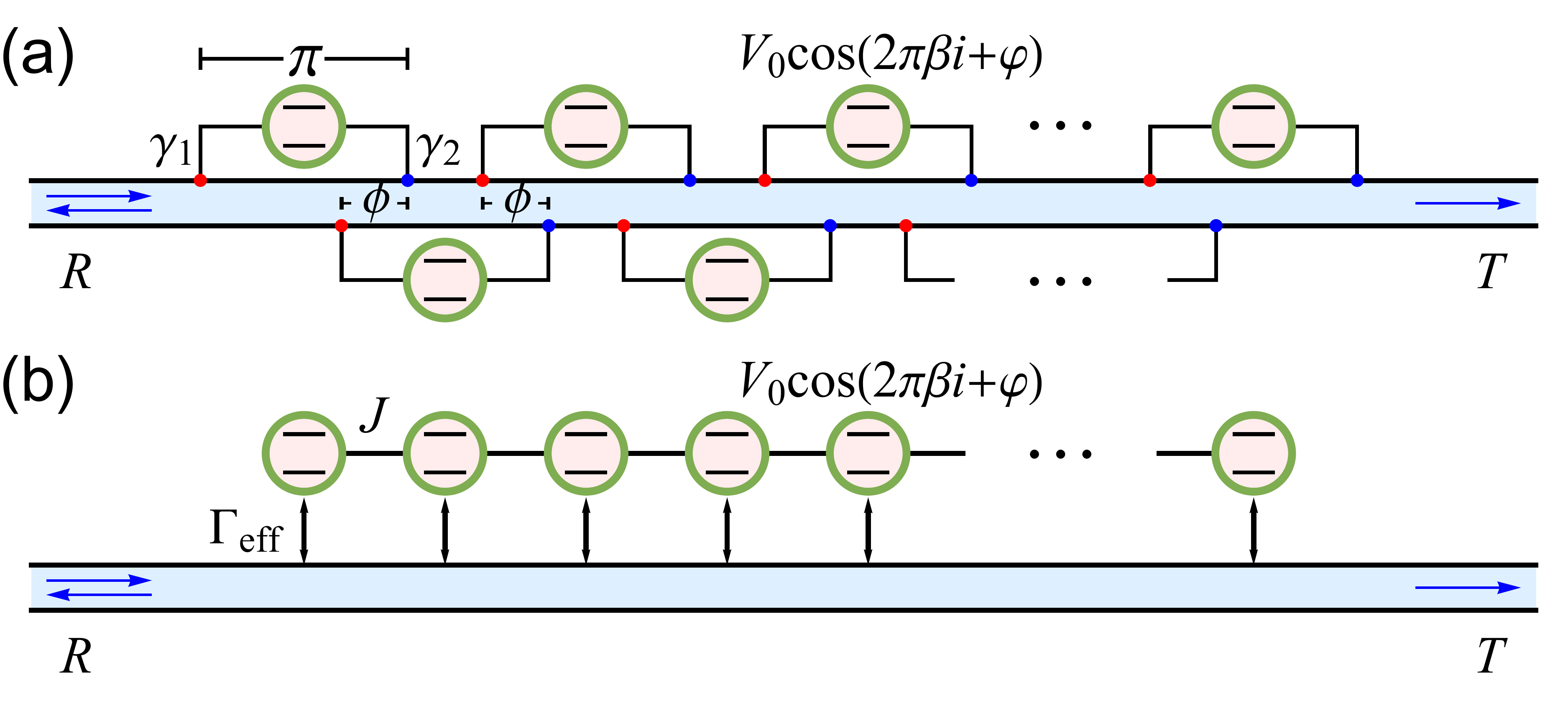}
	\caption{(a) Sketch of a setup with a braided giant-atom chain realizing an AAH chain with 
	decoherence-free nearest-neighbor couplings. (b) The effective system.
	}
	\label{SketchAAH}
\end{figure}
Here we focus on the wQED structures comprising $N$ two-level giant atoms with modulated frequencies 
$\omega_i= \omega_{\text{a}}+V_{0}\cos\left(2 \pi\beta{i}+\varphi\right)$ ($i=1,2,\cdots,N$). $V_{0}$ is the modulation amplitude, $\beta$ controls the 
periodicity of the modulation, and $\varphi$ is the modulation phase. Each atom is coupled to a 1D waveguide through two connection points and each 
pair of neighboring atoms is in a braided configuration, as shown in Fig.~\ref{SketchAAH}(a).
The implementation of this structure can be achieved through a variety of schemes, such as coupling superconducting qubits with multiple interdigital transducers to surface acoustic waves in a waveguide via piezoelectric effects~\cite{Andersson-NatPhy2019}, or coupling superconducting qubits capacitively to a meandering microwave transmission line at multiple points~\cite{Kannan-Nature2020}. And the frequency of each qubit can be modulated by a flux line.
The connection points are located at position $x_{im}$ ($i=1,2,\cdots,N$, and $m=1,2$), the
corresponding coupling strengths are $g_{i1}\equiv g_1$ and $g_{i2}\equiv g_2$, respectively.
The phase delays between the coupling points are set as $\phi=\theta_{i2}-\theta_{i+1,1}$ and
$\phi'=\theta_{i+1,1}-\theta_{i1}$ ($\phi\leq\phi'$), with $\theta_{im}=\omega_{\text{a}} x_{im}/v_{\text{g}}=k_{\text{a}}x_{im}$. 
$v_{\text{g}}$ is the group velocity of the photons in the
waveguide. Under the rotating-wave approximation, the Hamiltonian of the
system in the real space can be written as ($\hbar=1$)
\begin{equation}
	\begin{aligned}
	 \hat{H}=&\sum_{i=1}^{N}{\omega_i\hat{\sigma}_{i}^{+}\!\:\hat{\sigma}_{i}^{-}}+\int\mathrm{d}x\sum_s{\hat{c}_{s}^{\dagger}(x)\left(-\mathrm{i}l_sv_{\text{g}}\frac{\partial}{\partial x} \right)\hat{c}_s\left( x \right)}
	 \\
	&+\int\mathrm{d}x\sum_{s}\sum_{i=1}^{N}\sum_{m=1}^{2}{g_{m}}\delta \left( x-x_{im} \right) \left[ \hat{c}_{s}^{\dagger}\left( x \right) \hat{\sigma}_{i}^{-}+\text{H.c.} \right],
\end{aligned}
\label{AtomWGHamiltonian}
\end{equation}
where $s=\text{R}, \text{L}$ and $l_{\text{R},\text{L}} = \pm1$. $\hat{\sigma}_{i}^{+}$ ($\hat{\sigma}_{i}^{-}$)
is the raising (lowering) operator of
the atom $i$. $\hat{c}_{\text{R}}^{\dagger}(x)$ [$\hat{c}_{\text{R}}(x)$] and
 $\hat{c}_{\text{L}}^{\dagger}(x)$ [$\hat{c}_{\text{L}}(x)$] are the field operators of creating (annihilating)
the right- and left-propagating photons at position $ x$ in the
waveguide.

In the single-excitation subspace, the scattering
eigenstate of the system can be written as
\begin{equation}
|\Psi \rangle =\sum_s\int\mathrm{d}x\Phi_s(x)\hat{c}_{s}^{\dagger}(x)|\emptyset \rangle +\sum_{i=1}^{N}{f_i}\hat{\sigma}_{i}^{+}|\emptyset\rangle,
\label{IntEigenStat}
\end{equation}
where $|\emptyset\rangle$ is the vacuum state, which means that there are no photons in the waveguide, and meanwhile the atoms are in their ground states. $\Phi_{s}(x)$ ($s=\text{R}, \text{L}$) is the single-photon wave function in the $s$ mode. $f_{i}$ is the excitation amplitude of the atom $i$. Substituting Eq.~\eqref{IntEigenStat} into the eigen equation
\begin{equation}
\hat{H}|\Psi\rangle=\omega|\Psi\rangle
\end{equation}
yields the following equations of motion:
\begin{subequations}
\begin{equation}
\left(-\mathrm{i} v_{\text{g}} \frac{\partial}{\partial x}-\omega\right) \Phi_\text{R}(x)+\sum_{i=1}^{N}\sum_{m=1}^{2}g_{m}\delta\left(x-x_{im}\right) f_{i}=0,
\label{EoM1}
\end{equation}	
\begin{equation}
\left(\mathrm{i} v_{\text{g}} \frac{\partial}{\partial x}-\omega\right) \Phi_\text{L}(x)+\sum_{i=1}^{N}\sum_{m=1}^{2} g_{m} \delta\left(x-x_{im}\right) f_{i}=0,
\label{EoM2}
\end{equation}
\begin{equation}
\left(\omega_{i}-\omega\right) f_{i}+\sum_{s}\sum_{m=1}^{2} g_{m}\Phi_{s}\left(x_{i m}\right)=0.
\label{EoM3}
\end{equation}
\end{subequations}
\subsection{\label{ScatAmpAndEffH}Scattering amplitudes of photons and effective Hamiltonian of atom array}
We assume that a single photon with energy $\omega=v_{\text{g}} k$ is initially incident from the left, where $k$ is the wave vector of
the photon. Then $\Phi_\text{R}(x)$ and $\Phi_\text{L}(x)$ take the following \textit{ansatz}
\begin{subequations}
	\begin{equation}
\Phi_\text{R}(x)=e^{\mathrm{i}kx}\sum_{p=0}^{2N}{t_p}
\vartheta 
\left( x-x_p \right) \vartheta \left( x_{p+1}-x \right), 
\label{PhiR}
	\end{equation}	
	\begin{equation}
\Phi_\text{L}(x)=e^{-\mathrm{i}kx}\sum_{p=1}^{2N}{r_p}\vartheta \left( x-x_{p-1} \right) \vartheta \left( x_p-x \right). 
\label{PhiL}
	\end{equation}
\end{subequations}
Here the positions of coupling points from left to right are labeled as $x_1,x_2,\cdots,x_{2N}$. 
Moreover, $x_0=-\infty$ and $x_{2N+1}=+\infty$ are defined.
$t_p$ ($r_p$) is the transmission (reflection) 
amplitude of the $p$th coupling point. Particularly, $t_{2N}\equiv t$ ($r_{1}\equiv r$) is the transmission (reflection) amplitude
of the last (first) coupling point. $t_0=1$ is the amplitude of the incident field. $\vartheta (x)$ denotes the Heaviside step function.

Starting from the equations of motion \eqref{EoM1}-\eqref{EoM3} and the \textit{ansatz} 
\eqref{PhiR}-\eqref{PhiL}, and after some algebra \cite{Peng-PRA2023}, we can obtain the transmission and reflection amplitudes
\begin{subequations}
\begin{equation}
t=1-{\mathrm{i}}\mathbf{V}^{\dag}\left(\Delta\mathbf{I}-\mathbf{H} \right)^{-1}\mathbf{V},
\label{tGeneral}
\end{equation}
\begin{equation}
r=-{\mathrm{i}}\mathbf{V}^{\top}\left(\Delta\mathbf{I}-\mathbf{H} \right)^{-1}\mathbf{V}.
\label{rGeneral}
\end{equation}
\end{subequations}
The transmittance and the reflectance can be further defined as $T=|t|^2$ and $R=|r|^2$, respectively. 
In Eqs.~\eqref{tGeneral} and \eqref{rGeneral},  $\Delta=\omega-\omega_{\text{a}}$ is the photon-atom detuning, $\mathbf{I}$ is the identity matrix, and
$\mathbf{V}$ takes the form 
\begin{equation}
\mathbf{V}=(\mathcal{V}_1,\mathcal{V}_2,\cdots,\mathcal{V}_{N})^{\top},
\label{Vmatrix}
\end{equation}
with elements
\begin{equation} 
\mathcal{V}_{i}=\sum_{m=1}^{2}{\sqrt{\frac{\gamma _{m}}{2}}}e^{\mathrm{i}\theta_{im}}.
\end{equation}
Here the decay rate into the guided modes through the coupling point at $x_{im}$ is $\gamma_{m}=2g_{m}^2/v_{\text{g}}$. 
Note that we have assumed that the spacing between the connection points is small enough so that the phase-accumulated effects for detuned photons (the non-Markovian effects) can be neglected. Consequently, the wave vector $k$ in the definition of the phase factor has been replaced by $k_{\text{a}}$. $\mathbf{H}$ is the effective non-Hermitian Hamilton matrix of the atom array, with elements 
\begin{equation}
\mathcal{H}_{ij}=\left(\omega_i-\omega_{\text{a}}\right)\delta _{ij}-\frac{\mathrm{i}}{2}\sum_{m=1}^{2}\sum_{m'=1}^{2}{{\sqrt{\gamma _{m}\gamma _{m'}}}}e^{\mathrm{i}\left| \theta _{im}-\theta _{jm'}\right|},
\label{EffH}
\end{equation}
which summarize the coherent and dissipative atom-atom interactions mediated by the waveguide modes.
Moreover, this result indicates that the many-body model of the atomic chain can be designed by optimizing the layout of the connection points and the atom-waveguide coupling rates. Specifically, under the condition $\theta_{i2}-
\theta_{i1}=\phi+\phi'=(2n-1)\pi$ ($n\in\mathbb{Z}^{+}$), the Lamb shift of each atom vanishes, and
the corresponding effective Hamiltonian of the atom array reads 
\begin{eqnarray}
\hat H_{\text{eff}}&=&\sum_{i=1}^{N}\sum_{j=1}^{N}\mathcal{H}_{ij}\hat\sigma^{+}_i\hat\sigma^{-}_j
\nonumber
\\
&=&\sum_{i}\left[\left(\omega_{i}-\omega_{\text{a}}\right)-\frac{\mathrm{i}}{2}\Gamma _{\text{eff}}\right]\hat\sigma^{+}_i\hat\sigma^{-}_i
\nonumber
\\
&&+\sum_{i\neq j}\left(G_{ij}-\frac{\mathrm{i}}{2}\Gamma^{(\text{coll})}_{ij}\right)\hat\sigma^{+}_i\hat\sigma^{-}_j.
\label{Heff}
\end{eqnarray}
Here the effective decay of each atom is
\begin{equation}
\Gamma_{\text{eff}}=2\left(\gamma-\sqrt{\gamma^2-\delta^2}\right),
\label{Gammaeff}
\end{equation}
with $\gamma=(\gamma_1+\gamma_2)/2$ and $\delta=(\gamma_2-\gamma_1)/2$. And the exchange interaction and the collective decay between the $i$th and the $j$th ($i\neq j$) atoms take the form  
\begin{subequations}
\renewcommand{\arraystretch}{2}
\begin{equation}
G_{ij}=\left\{
\begin{array}{ll}
\gamma{\sin}\phi\equiv{J},&|i-j|=1
\\
\frac{(-1)^{|i-j|-1}}{2} \Gamma_{\text{eff}}\sin\left(|i-j|\phi\right),&|i-j|>1
\end{array}\right.,
\label{CoherentInt}
\end{equation}
\begin{equation}
\Gamma^{(\text{coll})}_{ij}=(-1)^{|i-j|-1}\Gamma_{\text{eff}}\cos[(i-j)\phi].
\label{CollDecay}
\end{equation}
\end{subequations}
\section{\label{SimulationAAH}Quantum simulation of AAH model through waveguide-QED systems with giant atoms}
\begin{figure*}[t]
	\centering
	\includegraphics[width=0.85\textwidth]{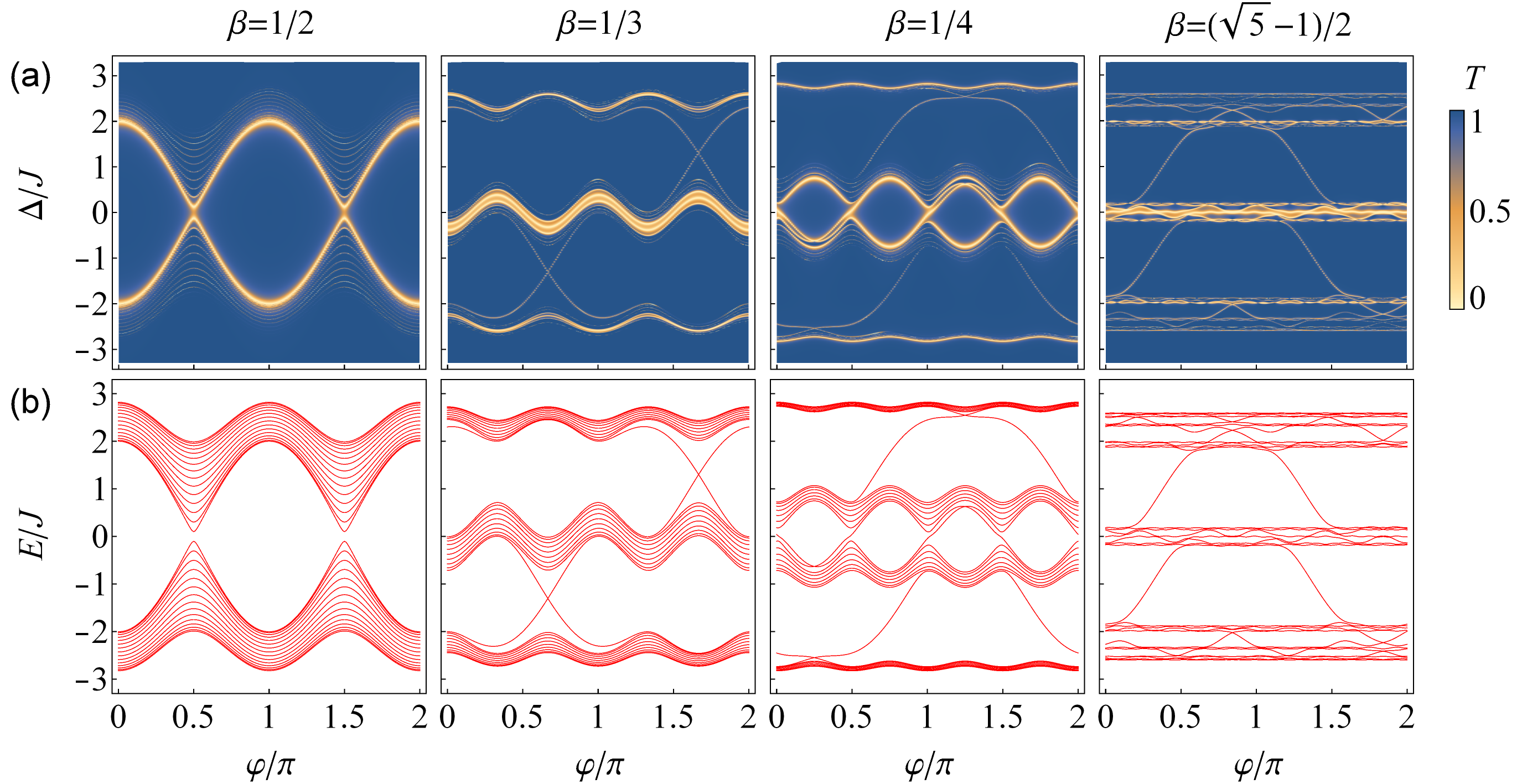}
	\caption{(a) Tansmitance for an AAH giant-atom chain with $N=30$ sites as a function of detuning $\Delta$ and modulation phase  
	$\varphi$ for different values of $\beta$. From left to right, $\beta$ is chosen as $1/2$, $1/3$, $1/4$, and $(\sqrt{5}-1)/2$, respectively. 
	The phase delay is selected to be $\phi=\pi/2$, resulting in $J=\gamma$.
	Other parameters are set as $\delta=0.1\gamma$ and $V_{0}=2J$. When $\beta$ is a rational number 
	$\beta=p/q$ (with $p$ and $q$ being coprime), 
	the corresponding spectrum is divided into $q$ bands 
	(see the first three panels). On the other hand, when $\beta$ is irrational, the spectrum become fractal 
	as a result of the quasiperiodicity of the system (see the last panel).
	(b) Energy spectra of an AAH model obtained from theoretical calculations for comparison 
	with (a).
	}
	\label{BandStructure}
\end{figure*}
\begin{figure*}[t]
	\centering
	\includegraphics[width=\textwidth]{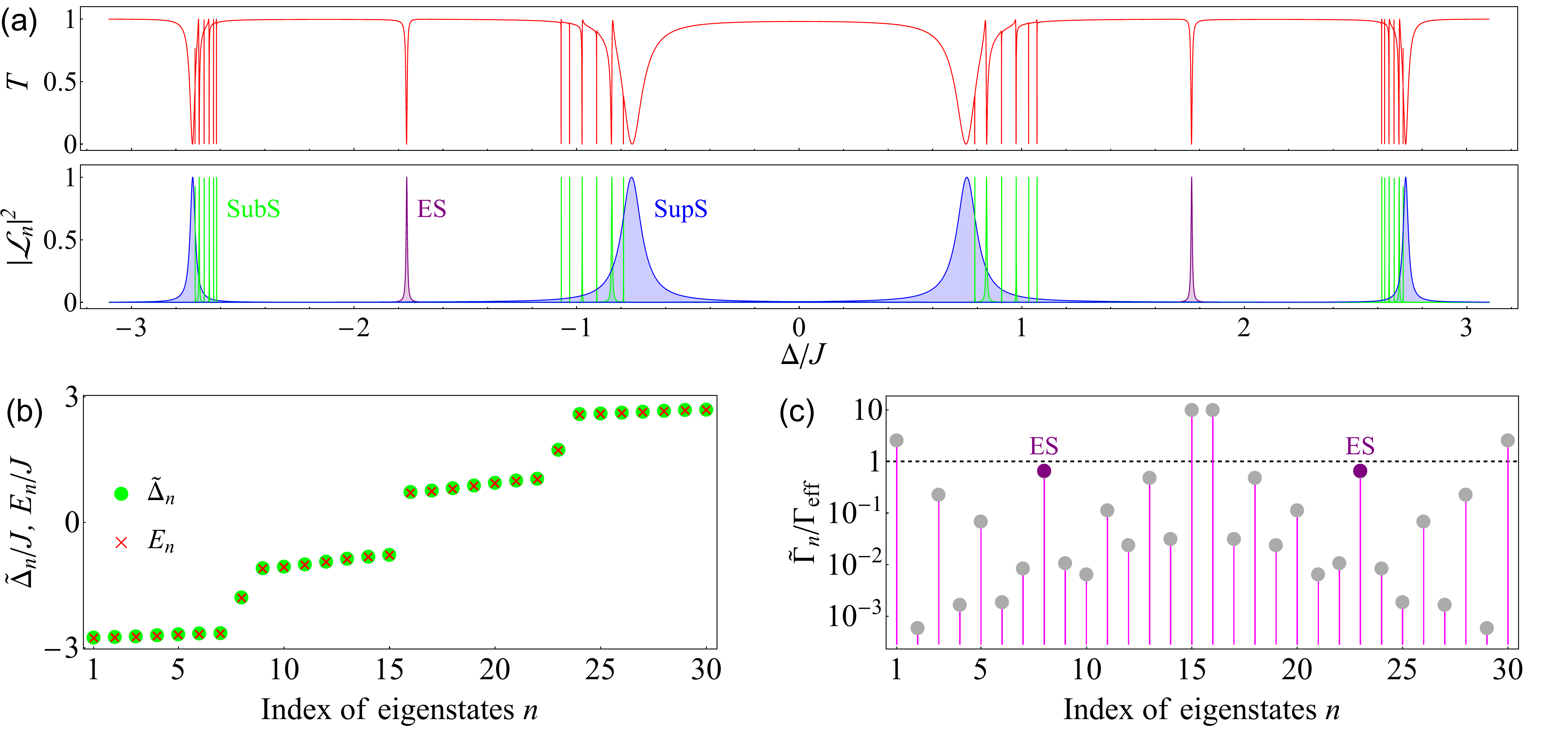}
	\caption{(a) Upper panel: transmission spectrum for an AAH giant-atom chain, with $\beta=1/4$
	  and $\varphi=3\pi/4$. Other parameters are the same as those used in Fig.~\ref{BandStructure} (a). Lower panel: the decomposed 
	  Lorentzian components for the spectrum shown in the upper panel. The resonances corresponding to the superradiant (subradiant) states in the 
	  band are labeled as SupS (SubS), with $\tilde{\Gamma}_n>\Gamma_{\text{eff}}$ ($\tilde{\Gamma}_n<\Gamma_{\text{eff}}$). The 
	  two edge states (ES) in the gaps exhibit a width that is approximately equal to $\Gamma_{\text{eff}}$. (b) The center frequency 
	  $\tilde{\Delta}_n$ (the disks) of each Lorentzian component and the eigenenergy $E_n$ (the crosses) given by the 
	  exact AAH model. (c) The linewidth $\tilde{\Gamma}_n$ of each Lorentzian component. The effective decay $\Gamma_{\text{eff}}$ of a 
	  single giant atom is labeled by the dashed line. 
	}
	\label{CollectiveModes}
\end{figure*}
\subsection{\label{AAHscheme} A scheme for constructing an AAH-type atomic chain using decoherence-free interactions}
Now we construct the AAH-type atomic chain based on the decoherence-free interactions between each pair of braided atoms, 
which is a unique property in giant-atom systems \cite{Kockum-PRL2018}.  To this end, we set $\gamma_1=\gamma_2=\gamma$ (i.e., $\delta=0$, 
together with $\theta_{i2}-\theta_{i1}=\phi+\phi'=(2n-1)\pi$, the condition 
for decoherence-free interactions between nearest neighbor atoms is satisfied).
From Eqs.~\eqref{Gammaeff} and \eqref{CollDecay}, it can be seen that the individual decays and the collective decays all vanish, with $\Gamma_{\text{eff}}=\Gamma^{(\text{coll})}_{ij}=0$.  And for exchange interactions, only the nearest-neighbor ones are nonzero, with $G_{ij}=\gamma\sin\phi\equiv J$ for $|i-j|=1$, and $G_{ij}=0$ for $|i-j|>1$ [see Eq.~\eqref{CoherentInt}]. 
Clearly, under these parameters the effective Hamiltonian \eqref{Heff} becomes the standard diagonal AAH Hamiltonian
\begin{eqnarray}
	\hat H_{\text{AAH}}&=&V_{0}\sum_{i=1}^{N}\cos\left(2 \pi\beta{i}+\varphi\right)\hat\sigma^{+}_{i}\hat\sigma^{-}_{i}
	\nonumber
	\\
	&&+J\sum_{i=1}^{N-1}\left(\hat\sigma^{+}_{i}\hat\sigma^{-}_{i+1}+\text{H.c.}\right).
	\label{Haah}
\end{eqnarray}

It should be emphasized that the phase delays $\phi$ and $\phi'$ can be manipulated by modulating the transition frequencies of the artificial atoms (e.g., 
through flux lines coupled to qubits in superconducting circuits~\cite{Kannan-Nature2020}). This 
manipulation can produce a range of values for the waveguide-induced decoherence-free interaction. Note that in experiments, the ratios $\zeta=\phi'/\phi$ are fixed in hardware by the relative lengths of the waveguide 
segments. Therefore, the condition $\phi+\phi'=(2n-1)\pi$ for decoherence-free interaction, under which the desired tight-binding atomic chain can be constructed, requires that the phase factor be tuned to $\phi=(2n-1)\pi/(\zeta+1)$, resulting in 
\begin{equation}
J=\gamma\sin\left[\frac{(2n-1)\pi}{\zeta+1}\right].
\end{equation}	
This allows us to get multiple values for the coupling strength $J\in[-\gamma,\gamma]$ (especially for relatively large $\zeta$), making the spectral width of the 1D chain highly tunable. Without loss of generality, in the following discussion we set the phase factor to be $\phi=\phi'=\pi/2$ (i.e., $\zeta=1,n=1$), so that the nearest-neighbor interaction strength reaches its maximum value $J=\gamma$.

Although a braided atomic chain under the above parameters can be exactly described by the diagonal AAH model, 
its states cannot be probed by photon scattering spectrum because all the atoms are decoupled from the waveguide.
Thus the accuracy of the simulated model must be balanced against the detectability of the atomic chain. This necessitates that the decay rates of the two connection points within each atom should be slightly different, with $\delta\ll\gamma$. Under this condition, 
each atom can obtain a tiny effective decay, which can be approximated as  
\begin{equation}
\begin{aligned}
\Gamma_{\text{eff}}\simeq\frac{\delta^2}{\gamma}
\end{aligned}.
\label{GammaeffAprox}
\end{equation}	
Thus the $i$th atom can interact with the photonic modes with a strength $\mathcal{V}_{i}\sim{\sqrt{{\Gamma _{\text{eff}}}/{2}}}$.
Furthermore, it can be readily demonstrated that for $\delta\ll\gamma$, the atom array can still be approximately described by the
AAH model. First for $\phi=\pi/2$ and $\delta\ll\gamma$, the nearest-neighbor interaction strength is still $J=\gamma$ [see Eq.~\eqref{CoherentInt}].
In addition, for non-nearest-neighbor interactions (with $|i-j|>1$), we have $|G_{ij}|/J=\Gamma_{\text{eff}}/(2\gamma)\simeq\delta^2/(2\gamma^2)$ for $|i-j|\in\mathbb{O}^{+}$ and $G_{ij}=0$ for $|i-j|\in\mathbb{E}^{+}$  [see Eq.~\eqref{CoherentInt}].
For collective decay rates, we have $\Gamma^{(\text{coll})}_{ij}=0$ for $|i-j|\in\mathbb{O}^{+}$
and $|\Gamma^{(\text{coll})}_{ij}|/J=\Gamma_{\text{eff}}/\gamma\simeq\delta^2/\gamma^2$ for $|i-j|\in\mathbb{E}^{+}$ [see Eq.~\eqref{CollDecay}].
These results demonstrate that, in comparison to nearest-neighbor coherent interactions, the non-nearest-neighbor coherent interactions and all 
dissipative interactions are either negligible or absent.
Thus the atomic chain can still be approximately described by the AAH Hamiltonian \eqref{Haah}, but each atom weakly couples to the 
waveguide modes. This implies that the braided atomic chain sketched in Fig.~\ref{SketchAAH}(a) is equivalent to the configuration shown in Fig.~\ref{SketchAAH}(b). The corresponding scattering amplitudes for incident photons can be calculated by using Eqs.~\eqref{tGeneral} and~\eqref{rGeneral}. 

Note that the above analysis is based on the assumption that the phase accumulation effects for detuned photons (the non-Markovian effects) are
negligible. If these effects are included, the phase factor $\phi$ should be replaced by $(1+\Delta/\omega_{\text{a}})\phi$~\cite{Cai-PRA2021}.  Thus, for 
$\phi=\pi/2$ considered here, the accumulated phase factor between the leftmost and
the rightmost connection points of the atomic chain is $(N+1)(1+\Delta/\omega_{\text{a}})\pi/2\simeq N(1+\Delta/\omega_{\text{a}})\pi/2$. 
Moreover, the detuning range we are interested in is approximately $\Delta\sim\gamma$.
Therefore, the extra phase $N\Delta\pi/(2\omega_{\text{a}})$ due to the effects of detuning  can be safely neglected if the condition
\begin{equation}
N\gamma\ll\omega_{\text{a}}
\label{MarkovCondition} 
\end{equation}
is satisfied. It is evident that for a 1D chain of tens to hundreds of atoms, this condition can be readily fulfilled within the currently available parameters of wQED systems~\cite{Sheremet-RMP2023,Kannan-Nature2020} (with $\omega_{\text{a}}/\gamma\sim10^3-10^4$). 

This approach to achieving many-body quantum simulations has the following features:

(i) The interactions between atoms are mediated by the waveguide modes, thereby obviating the necessity for direct couplings between atoms. In wQED 
with giant atoms, this kind of phase-dependent interactions are easy to modulate~\cite{Kannan-Nature2020,Kockum-PRL2018}. For instance, in our 
proposal, the tunability of the nearest-neighbor coupling strength for an AAH type tight-binding atomic chain can easily be implemented via a control line 
(e.g., a flux line~\cite{Kannan-Nature2020}) to modulate the atomic frequency. 
However, for small atoms, the creation of a tight-binding atomic chain with analogous tunability can only be achieved through the use of tunable coupling elements (usually in the form of superconducting qubits or superconducting quantum interference devices~\cite{Hime-Science2006,Niskanen-Science2007,Baust-PRB2015}), which requires more complex circuits.
 
(ii) In addition to inducing atom-atom interactions, the waveguide also acts as a measurement device, enabling the probing of the energy spectra and quantum states of the atomic chain through the photon scattering process. Thus 
additional readout devices coupled to the artificial atoms (e.g., resonators in superconducting circuits) are not required. 
This is particularly advantageous for atomic chains of large size, as it leads to substantial savings in hardware resources. 

(iii) In conventional spectroscopic technique \cite{Roushan-science2017,Shi-PRL2023} for detecting the spectral structure of a qubit array, it is necessary to perform the measurement on each atom in sequence. This is because measuring a single atom only provides partial information about the energy spectrum. In contrast, our method enables the complete energy spectrum to be revealed by \textit{a single scattering process}, as all atoms are coupled to the waveguide. Thus, our proposed measurement scheme may prove to be a highly efficient alternative to existing methodologies.
\subsection{\label{ProbeEnergyBand} Probing energy band structure through single-photon scattering spectra}
In Fig.~\ref{BandStructure}(a), we provide the transmittance $T$ for a lattice with $N=30$ as functions of detuning $\Delta$ and phase factor $\varphi$ for 
different values of $\beta$. In comparison to the energy spectra obtained through the diagonalization of the AAH Hamiltonian \eqref{Haah} [see 
Fig.~\ref{BandStructure}(b)], the transmission spectra displayed in Fig.~\ref{BandStructure}(a) can precisely extract the information of the band structures, with the transmission dips corresponding to the eigen energies of the atomic chain. It is known that the diagonal AAH model can be mapped onto a 2D Hofstadter lattice, where $\beta$ represents the number of magnetic flux quantum per unit cell and $\varphi$ denotes the momentum of an additional spatial dimension. In particular, when $\beta$ is a rational number $\beta=p/q$ (with $p$ and $q$ being coprime), 
the corresponding energy spectrum is divided into $q$ energy bands,  as shown by the first three panels in Figs.~\ref{BandStructure}(a) and \ref{BandStructure}(b). And for $\beta\neq 1/2$, the energy bands are with nontrivial topology, described by nonzero Chern numbers. Therefore, for a finite-sized system with boundaries considered here, energy levels corresponding to localized edge modes appear in the gaps. On the other hand,
in the case of irrational $\beta$ [see the last panels in Figs.~\ref{BandStructure}(a) and~\ref{BandStructure}(b)], the spectrum become fractal and form a Cantor set as a result of the quasiperiodicity of the system. The edge states can also be found in the gaps when the system is of finite size.

Moreover, Fig.~\ref{BandStructure}(a) shows that most transmission dips corresponding to the energy levels of the AAH atom array
are well resolved. To gain further insight into the formation of this type of spectral structure, we present in the upper panel of Fig.~\ref{CollectiveModes}(a) the transmission spectrum with $\beta=1/4$ and $\varphi=3\pi/4$, and decompose this spectrum into the superpositions of several Lorentzian-type amplitudes contributed by the collective excitations (see Appendix \ref{CollectiveModesDecomposition} for details):
\begin{equation}
		t=1+\sum_{n=1}^{N}{\eta}_n\mathcal{L}_{n},~~
		\mathcal{L}_{n}=\frac{\tilde{\Gamma}_n}{2\left(\Delta-\tilde{\Delta}_{n}+\mathrm{i}\frac{\tilde{\Gamma}_n}2\right)}.
		\label{LorentzianAmp}
\end{equation}	
Here $\tilde{\Delta}_n=\tilde{\omega}_n-\omega_\text{a}$ is the detuning between the $n$th collective mode and the atomic 
frequency. $\tilde{\Gamma}_{n}$ denotes the effective decay of the $n$th collective mode. ${\eta}_n$ determines the weight of the $n$th Lorentzian component. The obtained Lorentzian components are illustrated in the lower panel of Fig.~\ref{CollectiveModes}(a). 

It can be verified that the center frequencies of the Lorentzian components fit well with the energy levels given by the exact AAH model [see 
Fig.~\ref{CollectiveModes}(b)]. Additionally, as illustrated in Fig.~\ref{CollectiveModes}(c), only one of the bulk states in each band is superradiant (with $\tilde{\Gamma}_n>\Gamma_{\text{eff}}$), while the remainder are subradiant (with $\tilde{\Gamma}_n<\Gamma_{\text{eff}}$). 
The superposition and interference between these Lorentzian amplitudes give rise to the transmission spectrum depicted in the upper panel of Fig.~\ref{CollectiveModes}(a).  
Finally, the decay rate of the two edge states in the gaps is approximately equal to the effective decay rate $\Gamma_{\text{eff}}$ of a single atom 
[see Fig.~\ref{CollectiveModes}(c)], given that in these states, only the atom at one end is highly excited.  
It should be noted that the assumptions of $\delta\ll\gamma$ (resulting in $\Gamma_{\text{eff}}\simeq\delta^2/\gamma$) 
and $\phi=\pi/2$ (resulting in $J=\gamma$)
have been made to ensure that the atom array can be approximately 
described by the AAH model. Consequently, the relation 
$N\Gamma_{\text{eff}}/J\simeq N\delta^2/\gamma^2\ll1$ can be easily satisfied for not so large 
$N$. Namely, the widths of the resonances (the majority of which are smaller than the value of $\Gamma_{\text{eff}}$) are considerably less than the spectrum width (which is of the order of $J$) divided by $N$.
This property offers a high resolution for the transmission dips associated with the energy levels of the atomic chain, thereby facilitating precise probing of the band structures, as shown in Fig.~\ref{CollectiveModes}(a). The following section will further illustrate the potential for simulating the Hofstadter butterfly spectrum by exploiting this advantage of giant-atom wQED.
\begin{figure}[t]
	\centering
	\includegraphics[width=0.5\textwidth]{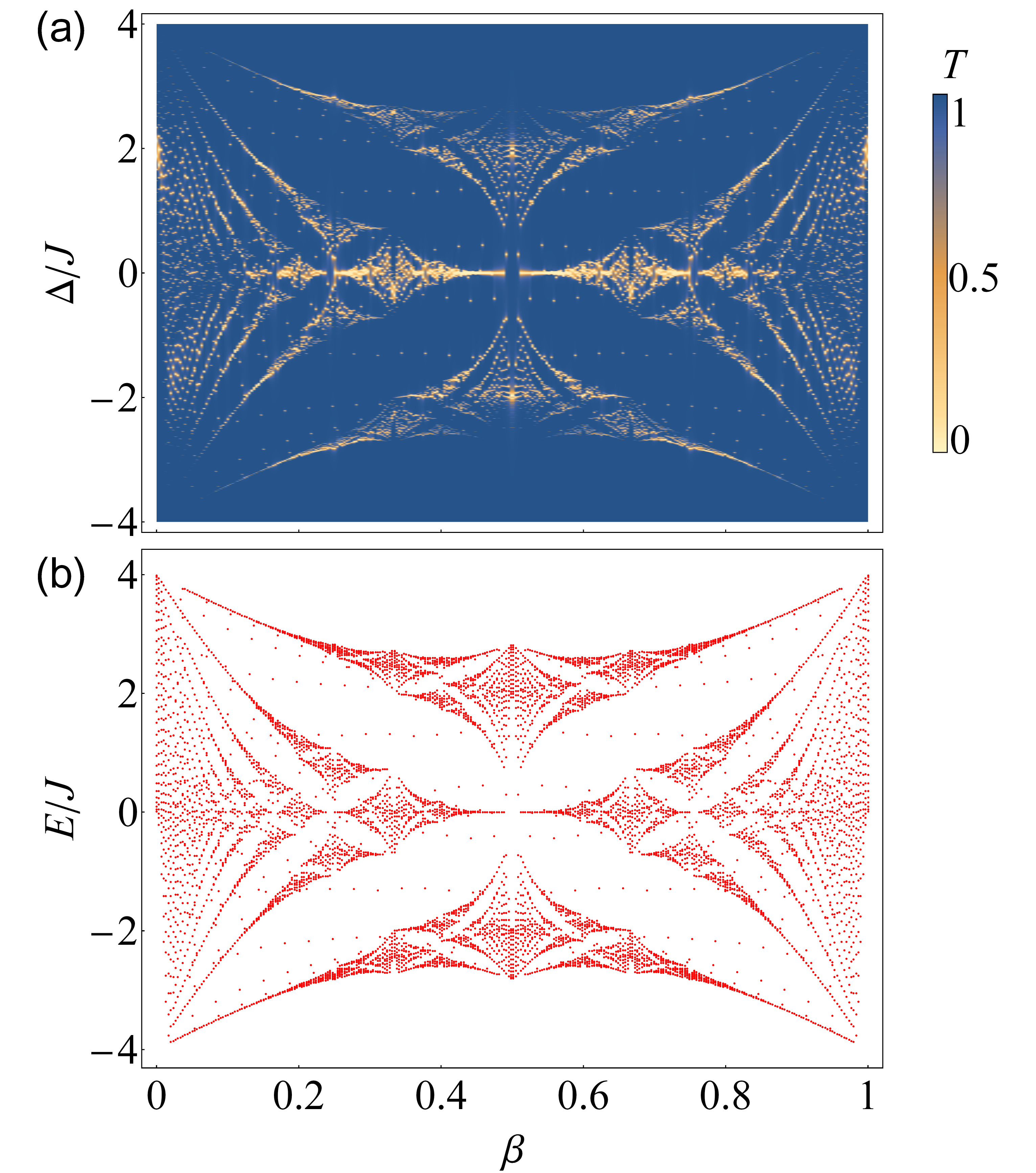}
	\caption{(a) Quantum simulation of Hofstadter butterfly using scattering spectra of a giant-atom wQED setup. The modulation phase is set as $\varphi=0$. The modulation frequency‌ $\beta$ is varied from $0$ to $1$, generating $299$ distinct instances of the transmission spectrum. Other parameters are the same as those used in Fig.~\ref{BandStructure}(a). (b) The theoretical Hofstadter butterfly energy spectrum obtained from the AAH Hamiltonian.
	}
	\label{Butterfly}
\end{figure}
\begin{figure*}[t]
	\centering
	\includegraphics[width=\textwidth]{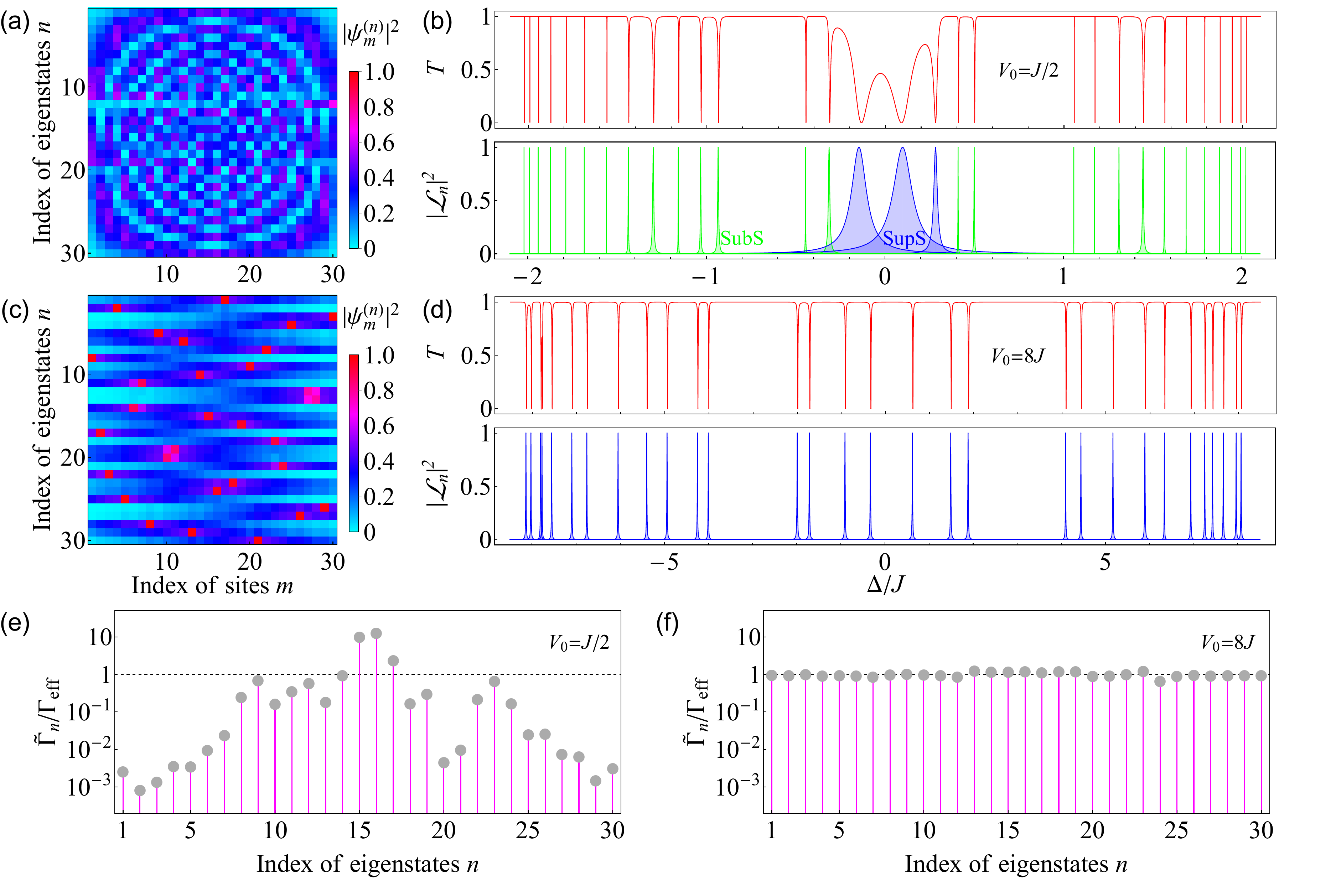}
	\caption{(a) The square of the modulus of the wave functions of AAH model with $N=30$, $V_0=J/2$, $\beta=(\sqrt{5}-1)/2$, and 
	$\varphi=0$. (b) The transmission
	 spectrum (upper panel) and corresponding Lorentzian decompositions (lower panel) for a wQED system with an AAH-type giant-atom
	array. The phase delay is selected to be $\phi=\pi/2$, resulting in $J=\gamma$. The quantity $\delta$ is chosen as $\delta=0.1\gamma$. The size of the array and the on-site modulations are the same
	as in (a). In the lower panel, the resonances corresponding to the superradiant (subradiant) states are labeled as SupS (SubS).
	(c) and (d) Similar to (a) and (b), but with $V_0=8J$.     
	(e) snd (f) The decay rate $\tilde{\Gamma}_n$ of each Lorentzian component for $V_0=J/2$ and $V_0=8J$. The effective decay $\Gamma_{\text{eff}}$ of a 
	single giant atom is labeled by the dashed line.} 
	\label{Localization}
\end{figure*}
\begin{figure}[t]
	\centering
	\includegraphics[width=0.5\textwidth]{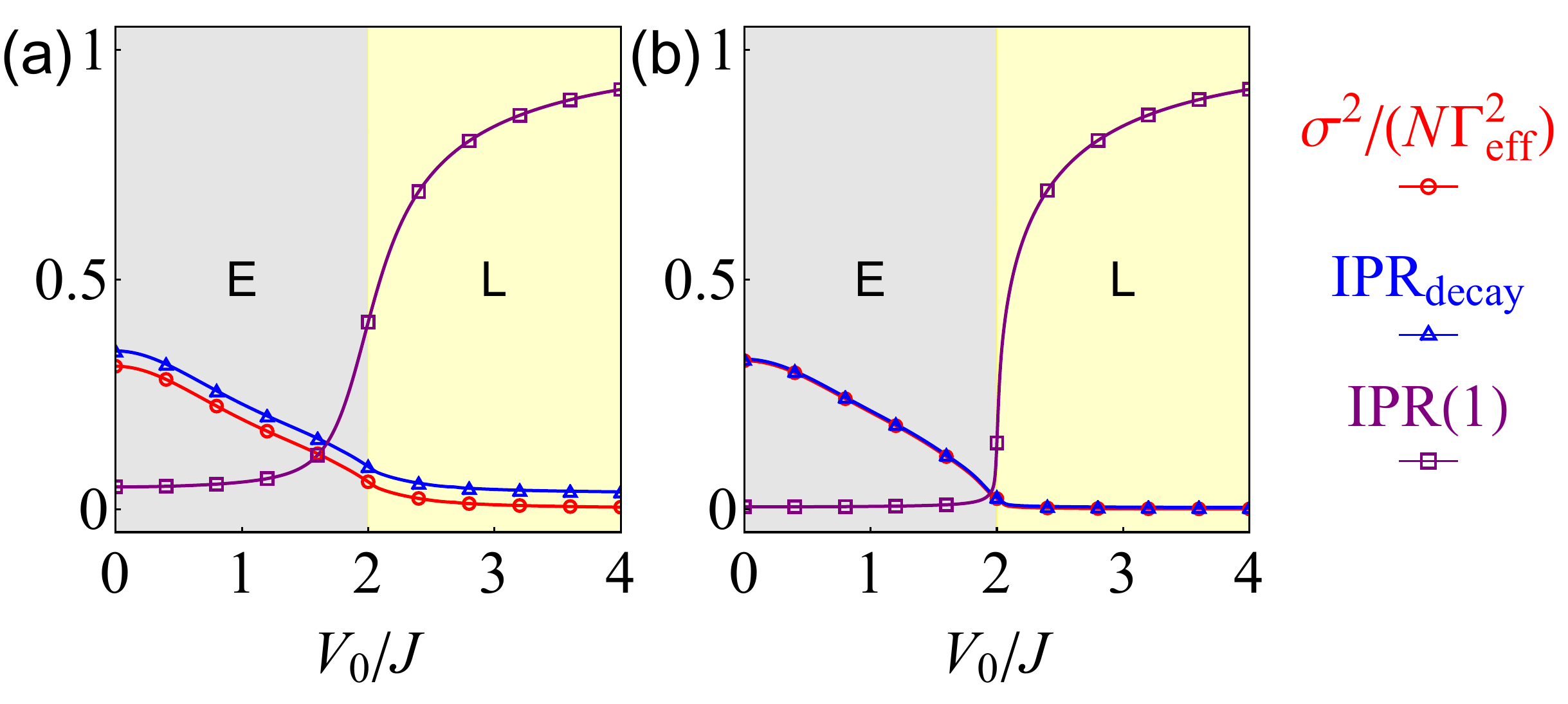}
	\caption{$\sigma^2$ and $\text{IPR}_{\text{decay}}$ as functions of $V_{0}$, with
	(a) $N=30$, $\delta=0.1\gamma$, and (b) $N=300$, $\delta=0.01\gamma$.  
	Other parameters are the same as those used in 
	Fig.~\ref{Localization}. The extended (E) and the localized (L) phases are well characterized by these quantities. In panels (a) and (b), the inverse participation ratio $\text{IPR}(1)$ (Here 1 is used to denote the ground state)
	of the standard AAH chain is also provided for comparison.} 
	\label{IPR}
\end{figure}
\subsection{\label{ProbeButterfly}Reconstructing the fractal structure of the Hofstadter butterfly spectrum}

It is well known that the problem of Bloch electrons on a 2D lattice subject to a perpendicularly applied magnetic field is described by the Hofstadter model. When the magnetic field reaches a strength that allows one magnetic flux quantum per unit cell to pass through, 
the corresponding energy spectrum versus the dimensionless magnetic flux form a famous fractal structure known as the Hofstadter butterfly \cite{Hofstadter-PRB1976}. 
Nevertheless, a magnetic field of approximately $10^4$T is necessary to observe the Hofstadter butterfly in typical crystals, which is beyond the capabilities of current experimental techniques. One potential solution to this challenge is the use of superlattices with a larger lattice constant, which could reduce the requisite strength of magnetic fields \cite{Dean-Nature2013,Ponomarenko-Nature2013,Hunt-science2013}. On the other hand, the 2D Hofstadter model can be exactly mapped to the 1D diagonal AAH model. Consequently, the Hofstadter butterfly spectrum has also been successfully simulated in 1D AAH-type chains based on superconducting circuits \cite{Shi-PRL2023,Roushan-science2017}. In this approach, the modulation frequency $\beta$ is used to analogize the required magnetic flux, thereby eliminating the need for a complex 2D lattice and an external magnetic field. 

As previously demonstrated in Sec.~\ref{ProbeEnergyBand}, in the AAH-type qubit array based on giant-atom wQED, the majority of 
the resonance dips (corresponding to the energy levels of the AAH model) in the transmission spectrum can be well resolved. 
Therefore, the transmission spectrum of this system can be effectively utilized to reconstruct the complex and fine fractal structure of the Hofstadter 
butterfly spectrum with a high degree of fidelity, as shown in Fig.~\ref{Butterfly}(a). In this panel, the number of giant atoms is set to $N=30$.  
The reduced magnetic flux‌ $\beta$ is varied from $0$ to $1$, generating $299$ instances of transmission spectra, which exhibit 
clear fractal features and a shape that is reminiscent of a butterfly. 
Figure~\ref{Butterfly}(b) depicts the Hofstadter butterfly energy spectrum obtained from the theoretical model,
which exhibits a high degree of agreement with the result
 in Fig.~\ref{Butterfly}(a). Thus, it appears that an alternative or potentially more efficient approach 
to simulating the Hofstadter butterfly spectrum is feasible.
\subsection{\label{LocalizationTransition}Probing the localization transition}
The localization transition \cite{Aubry-IPS1980,Simon-AAM1982,Jitomirskaya-JSTOR1999} is another interesting topic in the incommensurate diagonal AAH model.
When the on-site potential of the diagonalized AAH model is made quasiperiodic by setting $\beta$ to an irrational number, the model exhibits a localization transition, where all bulk eigenstates are extended for $V_0<2J$, and localized for $V_0>2J$.  
Here we will present an intriguing correlation between the localization transition of the atomic chain and the photon transport behavior observed in the waveguide.   

To analyze the relationship between the localization transition and the scattering spectrum, we let the value of the modulation frequency be
irrational, with $\beta=(\sqrt{5}-1)/2$, and present in Fig.~\ref{Localization}(a) the square of the modulus of the wave functions $\psi_m^{(n)}$ 
(where $m$ and $n$ are used to index the sites and eigenstates, respectively) of the diagonal AAH model for $V_0=J/2$, which corresponds to the extended phase. The resulting transmission spectrum and the corresponding Lorentzian decompositions are illustrated in Fig.~\ref{Localization}(b). 
These collective modes with extended feature comprise a small number of superradiant states (with $\tilde{\Gamma}_n>\Gamma_{\text{eff}}$) and a significant proportion of subradiant states (with $\tilde{\Gamma}_n<\Gamma_{\text{eff}}$), exhibiting a wide range of linewidths, as illustrated in the lower panel of Fig.~\ref{Localization}(b).  
While for $V_0=8J$, the situation is quite different. As shown in Fig.~\ref{Localization}(c), the wave functions of the atomic chain
are highly localized. In this case, the decay of each collective mode to the waveguide is primarily attributable to the most excited atom, which has an excitation probability close to one. Therefore, the width of each resonance dip in the transmission spectrum [see the upper panel of Fig.~\ref{Localization}(d)] and the width of each collective excitation amplitude [see the lower panel of Fig.~\ref{Localization}(d)] are approximately equal to the effective decay rate $\Gamma_{\text{eff}}$ of a single atom. Namely, all collective states exhibit neither a markedly superradiant nor a distinctly subradiant character.

A more straightforward illustration of the decay rates of the collective modes as a function of index $n$ in the extended and localized phases can be found in Figs.~\ref{Localization}(e) and \ref{Localization}(f), respectively.
It can be observed that there exists an interesting correspondence
between the extended (localized) phase of the atom array and the ``localization" (``delocalization") of the decay rates of its collective modes.    

We now seek ways to quantitatively characterize this correspondence. 
Note that the degree of localization of the $n$th eigenstate of the AAH chain can be described by the inverse participation ratio (IPR):
\begin{equation} 
\text{IPR}(n)=\sum_{m}\left|\psi^{(n)}_m\right|^{4},
\label{IPRWF}
\end{equation}
which vanishes for extended states [see the curves marked with squares in Figs.~\ref{IPR}(a) and \ref{IPR}(b)]. As for the aspect of the collective modes,  the mean value of their decay rates is equal to
the decay rate of a single atom, with  
$(\sum_{n}\tilde\Gamma_{n})/N=\Gamma_{\text{eff}}$.
It is therefore reasonable to conclude that the variance 
\begin{equation} 
\sigma^2=\frac{1}{N}\sum_{n}\left(\tilde\Gamma_{n}-\Gamma_{\text{eff}}\right)^2
\label{variance}
\end{equation}
may be employed as a means of measuring
the concentration of the decay-rate distribution (vs index $n$).
It is evident that the variance $\sigma^2\simeq 0$ when $V_0>2J$, and $\sigma^2 > 0$ when $V_0<2J$ [see the
curves marked with circles in Fig.~\ref{IPR}(a) and \ref{IPR}(b)], indicating the localized and 
extended phases, respectively. Note that when $V_0\to 0$, the order of magnitude of $\sigma^2$ is $N\Gamma_{\text{eff}}^2$ for 
an array with $N\gg 1$. This is due to 
the fact that in this regime, the decay rates of the superradiant modes,
which are a few in number and of 
the order of $N\Gamma_{\text{eff}}$, contribute mainly to the variance.    
One can further define a ``normalized" decay rate $\tilde{\Gamma}_{\text{norm},n}=\tilde\Gamma_{n}/(N\Gamma_{\text{eff}})$, with $\sum_{n}\tilde{\Gamma}_{\text{norm},n}=1$ (which is analogous to the normalization condition $\sum_{m}|\psi^{(n)}_m|^{2}=1$ of the wave functions). Therefore, it is possible to emulate the IPR for the wave functions [see Eq.~\eqref{IPRWF}] to define the following IPR describing the distribution of the decay rates 
\begin{equation} 
\text{IPR}_{\text{decay}}=\sum_{n}\tilde{\Gamma}^2_{\text{norm},n}.
\label{IPRdecay}
\end{equation}
The inverse of this quantity indicates the number of the collective modes that can decay appreciably into the waveguide.
As shown in Fig.~\ref{IPR}, both $\sigma^2$ and $\text{IPR}_{\text{decay}}$ can be used to quantify the localization transition. 
Moreover, in the thermodynamic limit $N\to\infty$, we have $\sigma^2=N\Gamma^2_{\text{eff}}\times\text{IPR}_{\text{decay}}$.
This is consistent with the results (see the curves marked with circles and triangles) in Fig.~\ref{IPR}(b), where a relatively large value of $N=300$ is chosen.
Besides, a comparison of Figs.~\ref{IPR}(a) and \ref{IPR}(b) illustrates that a finite-size system comprising a greater number of sites is more effective at exhibiting characteristics associated with quantum phase transitions.  

In summary, the many-body state of the atomic chain is profoundly imprinted on the information contained within the single-photon
scattering spectrum, making the localization transition easy to detect in wQED systems. 
\subsection{\label{DissipationUnguided}Influence of spontaneous emission to channels other than the waveguide continuum}
\begin{figure}[t]
	\centering
	\includegraphics[width=0.5\textwidth]{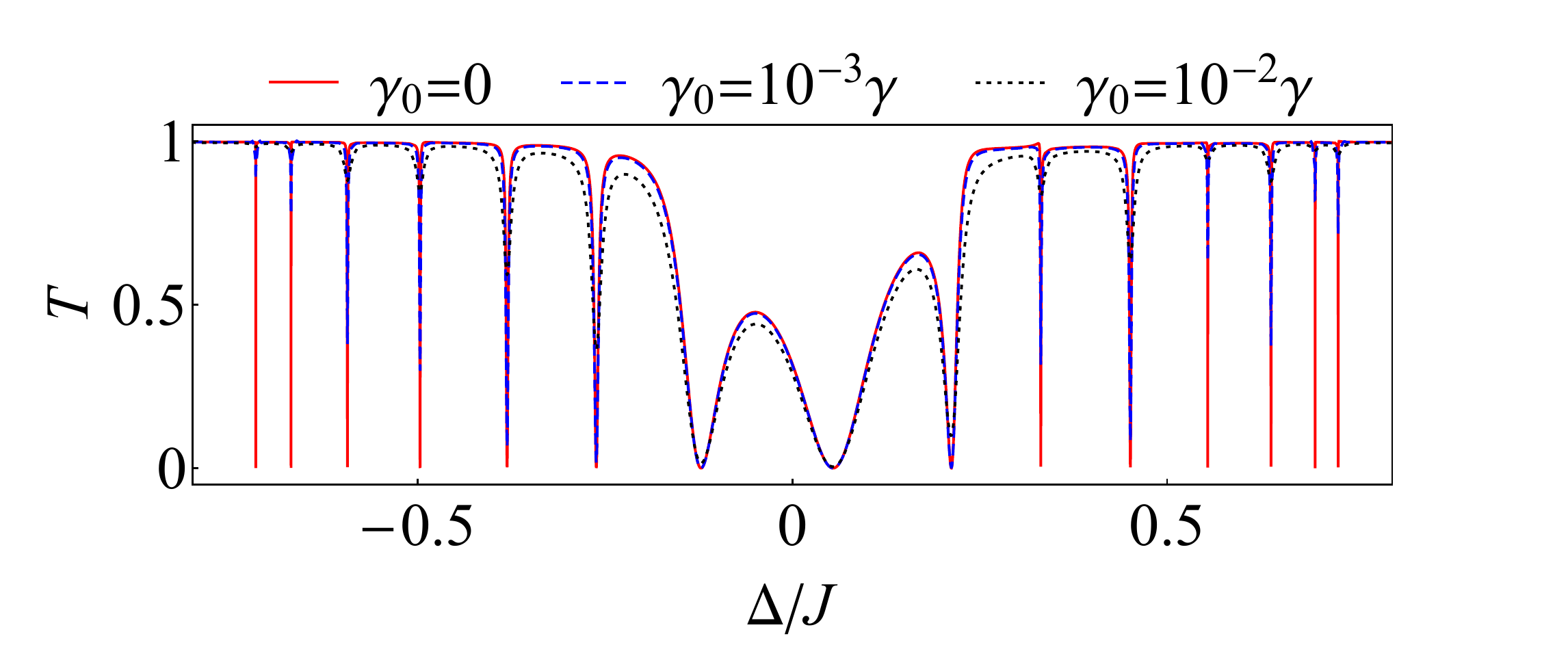}
	\caption{Transmission coefficients as functions for different values of $\gamma_0$. The modulation frequency and the modulation phase
	are chosen as $\beta=1/4$ and $\varphi=0$. Other parameters are the same as those used in Fig.~\ref{BandStructure}(a). The range of $\Delta$ is selected to display the spectrum corresponding to the center band.}
	\label{InfluenceLoss}
\end{figure}

In practice, the unavoidable photon loss resulting from the coupling of the system with the surrounding environmental degrees of freedom should be taken into account. Assuming that all atoms have the same photon loss rate $\gamma_0$, the Hamiltonian~\eqref{AtomWGHamiltonian} of the system can be rewritten as the following
non-Hermitian one $\hat{H}'=\hat{H}-(\mathrm{i}\gamma_{0}/2)\sum_{i=1}^{N}{\hat{\sigma}_{i}^{+}\!\:\hat{\sigma}_{i}^{-}}$. 
The corresponding scattering amplitudes can be obtained by replacing the term $\mathbf{H}$ in Eqs.~\eqref{tGeneral} and \eqref{rGeneral} by
$\mathbf{H}-{\mathrm{i}\gamma_0}\mathbf{I}/2$. And the effective decay of the $n$th collective mode
 becomes $\tilde\Gamma'_n=\tilde\Gamma_n+\gamma_0$. Obviously, the collective modes with small coupling 
rates to the waveguide (with $\tilde\Gamma_n\ll\gamma_0$) are hardly excited due to their extremely low coupling efficiency to the waveguide. Thus it can be observed from the transmission spectrum that the dips with narrower widths (corresponding to the subradiant states) are more significantly suppressed when a photon-loss rate $\gamma_0$ is included, as shown in Fig.~\ref{InfluenceLoss}. 
For an atom array with dissipation rate of $\gamma_0=10^{-3}\gamma$ (which can be achieved in current wQED setups based on superconducting quantum circuits \cite{Mirhosseini-Natrue2019}), all the collective modes of the atom array can still be probed by the transmission spectrum, including the
extremely subradiant ones near the band edges, as shown by the blue dashed line in Fig.~\ref{InfluenceLoss}.

Now we discuss the influence of photon loss on probing the many-body localization transition. 
When a photon loss $\gamma_0\ll\Gamma_{\text{eff}}$ is included, all the line widths (denote as $\tilde\Gamma'_{n}=\tilde\Gamma_{n}+\gamma_0$), as 
well as their average (denote as $\Gamma'_{\text{eff}}=\sum_{n}\tilde\Gamma’_{n}/N=\Gamma_{\text{eff}}+\gamma_0$, 
which is equal to the total decay rate of a single atom), undergo an identical increment $\gamma_0$. Thus, the width variance remains unchanged even after 
accounting for photon loss, i.e., $\sigma'^2=\sum_{n}(\tilde\Gamma'_{n}-\Gamma'_{\text{eff}})^2/N=\sigma^2$ [see Eq.~\eqref{variance}] is fulfilled.
Therefore, as long as $\gamma_0\ll\Gamma_{\text{eff}}$ is satisfied so that all the widths $\tilde\Gamma'_{n}$ of the resonance dips can be measured, the corresponding variance $\sigma'^2$ can still accurately indicate whether the system is in the localized or extended phase. On the other hand, 
using these measured $\tilde\Gamma'_{n}$, the IPR describing the distribution of the decay rates 
$\text{IPR}'_{\text{decay}}=\sum_{n}\tilde{\Gamma}'^2_{\text{norm},n}$ can also be constructed [similar to $\text{IPR}_{\text{decay}}$ defined  in Eq~\eqref{IPRdecay}], with $\tilde{\Gamma}'_{\text{norm},n}=\tilde\Gamma'_{n}/
(N\Gamma'_{\text{eff}})$ and $\sum_{n}\tilde{\Gamma}'_{\text{norm},n}=1$. It can be easily proved that
in the thermodynamic limit, the relation $\text{IPR}'_{\text{decay}}\simeq (1-2\gamma_0/\Gamma_{\text{eff}})\text{IPR}_{\text{decay}}$ is satisfied.  
Clearly, $\text{IPR}'\simeq 0$ ($\text{IPR}'> 0$) is satisfied when 
the system is in the localized (extended) phase. Therefore, the quantity $\text{IPR}'_{\text{decay}}$ can also characterize the many-body quantum phase transition in the photon-loss case. 
\section{\label{conclusion}conclusion}
In summary, we present an experimentally feasible scheme to simulate and probe the AAH model by utilizing 
giant-atom wQED systems. In our proposal, the coupling between a pair of neighboring atoms is generated by the nearly decoherence-free interaction mediated by the waveguide modes, thus eliminating the need for direct couplings between atoms. In addition to mediating interactions between atoms, the photonic modes in the waveguide are also used to detect the energy spectrum of the atom array. Moreover, our method allows the energy spectrum of the atomic chain to be revealed by a single scattering process. As a result, the proposed system is characterized by a simple structure and high detection efficiency.
In addition, the optimized parameters for achieving high-precision quantum simulation are analyzed to demonstrate the feasibility of the scheme under existing experimental techniques. 

As specific examples, we first demonstrate the feasibility of precisely reconstructing the complex and fine fractal structure of the Hofstadter butterfly by utilizing the transmission spectrum. Subsequently, we show that in the case of incommensurate diagonal AAH model, the statistical distribution of the linewidths of the collective modes can be employed as a means of quantifying the localization transition, thereby facilitating the detection of this transition. 

Our method is also applicable to the construction and detection of other types of 1D atomic chains, thus paving the way for quantum simulations of more complex many-body states based on giant-atom wQED. Furthermore, the results of this study suggest that wQED with giant atoms is an ideal platform for investigating light-matter interactions between a coupled qubit (or spin) chain and its photonic environment.
\begin{acknowledgments}
This work was supported by the National Natural Science Foundation of China (NSFC) under Grants No. 61871333.
\end{acknowledgments}
\appendix
\section{\label{CollectiveModesDecomposition}Scattering amplitudes expressed in terms of collective modes}
To better understand the physics of the scattering process, we rewrite the scattering amplitudes \eqref{tGeneral} and \eqref{rGeneral} in terms of 
collective modes of the atom array: 
\begin{subequations}
	\begin{equation}
		t =1-\mathrm{i}\sum_{n=1}^{N}\frac{\big(\mathbf{V}^{\dagger}{\mathbf{U}^{\mathscr{R}}_{n}}\big)\big({\mathbf{U}^{\mathscr{L}}_{n}}^\dagger\mathbf{V}\big)}{\Delta-\lambda_n},
		\label{tDecompose1}
	\end{equation}
	\begin{equation}
		r =-\mathrm{i}\sum_{n=1}^{N}\frac{\big(\mathbf{V}^{\top}{\mathbf{U}^{\mathscr{R}}_{n}}\big)\big({\mathbf{U}^{\mathscr{L}}_{n}}^\dagger\mathbf{V}\big)}{\Delta-\lambda_n},
		\label{rDecompose1}
	\end{equation}	
\end{subequations}
where $\mathbf{U}^{\mathscr{R}}_{n}$ and $\mathbf{U}^{\mathscr{L}}_{n}$ are the right and left eigenvectors of the non-Hermitian Hamilton matrix $\mathbf{H}$, and $\lambda_n$ is the corresponding complex eigenvalues, satisfying 
$\mathbf{H}\mathbf{U}^{\mathscr{R}}_{n}=\lambda_{n}\mathbf{U}^{\mathscr{R}}_{n}$, 
$\mathbf{H}^{\dag}\mathbf{U}^{\mathscr{L}}_{n}=\lambda^{*}_{n}\mathbf{U}^{\mathscr{L}}_{n}$,
and ${\mathbf{U}^{\mathscr{L}}_{n}}^\dag\mathbf{U}^{\mathscr{R}}_{n'}={\mathbf{U}^{\mathscr{R}}_{n}}^\dag\mathbf{U}^{\mathscr{L}}_{n'}=\delta_{nn'}$ \cite{Brody-JPA2013}.
The numerators in Eqs.~\eqref{tDecompose1} and \eqref{rDecompose1} represent the overlap degree of the $n$th collective state of atomic chain and the propagating photon modes.

Equations \eqref{tDecompose1} and \eqref{rDecompose1} show that the scattering spectrum can be regarded as the result of interference between different scattering channels, which are provided by the corresponding collective modes.
Under the Markovian approximation, the input $\mathbf{V}$, the Hamiltonian $\mathbf{H}$, and consequently the eigenvalues $\lambda_n$ and eigenvectors $\mathbf{U}^{\mathscr{R,L}}_{n}$ of
$\mathbf{H}$, are independent of the frequency of the photons. Accordingly, the eigen frequency (with respect to the reference frequency $\omega_\text{a}$) and the effective decay of the $n$th collective mode can be defined as $\tilde{\Delta}_n=\mathrm{Re}(\lambda_n)$ and $\tilde{\Gamma}_{n} = -2\mathrm{Im}(\lambda_n)$, respectively. Further, we can define the weight factors $\eta_n =-2\mathrm{i}[{\big(\mathbf{V}^{\dagger}{\mathbf{U}^{\mathscr{R}}_{n}}\big)\big({\mathbf{U}^{\mathscr{L}}_{n}}^\dagger\mathbf{V}\big)}]/\tilde{\Gamma}_{n}$ and $\xi_n=-2\mathrm{i}[{\big(\mathbf{V}^{\top}{\mathbf{U}^{\mathscr{R}}_{n}}\big)\big({\mathbf{U}^{\mathscr{L}}_{n}}^\dagger\mathbf{V}\big)}]/\tilde{\Gamma}_{n}$. Thus the transmission and reflection amplitudes can be written as
\begin{subequations}
	\begin{equation}
		t=1+\sum_{n=1}^N\frac{ {\eta_n}{\tilde{\Gamma}_n}}{2\left(\Delta-\tilde{\Delta}_n+\mathrm{i}\frac{\tilde{\Gamma}_n}{2}\right)},
		\label{tDecompose2}
	\end{equation}	
	\begin{equation}
		r=\sum_{n=1}^N\frac{{{\xi}_n}{\tilde{\Gamma}_n}}{2\left(\Delta-\tilde{\Delta}_{n}+\mathrm{i}\frac{\tilde{\Gamma}_n}2\right)}.
		\label{rDecompose2}
	\end{equation}	
\end{subequations}
The above equations illustrate the superposition of several Lorentzian-type amplitudes contributed by the collective excitations, which is very helpful for us to analyze the scattering spectra for multi-atom wQED.
%
%
\bibliography{MS-Dec23-2024YCL}
\end{document}